\documentclass{PoS}

\title{Theory of hard probes in PbPb collisions}

\ShortTitle{Jet substructure and cross section in heavy ion collisions using SCET}

\author{\speaker{Yang-Ting Chien}\thanks{The speaker would like to thank Konrad Tywoniuk for the warm invitation and the stimulating discussions, and Lund University for the hospitality during the conference.}\\
        Theoretical Division, T-2, Los Alamos National Laboratory, Los Alamos, NM 87545, U.S.A.\\
        E-mail: \email{ytchien@lanl.gov}}

\abstract{The jet quenching phenomenon in heavy ion collisions provides a strong evidence of the modification of parton shower in the quark-gluon plasma. This contribution focuses on the hard probes of QGP using jets and summarizes the new theoretical progress of jet substructure modification studies using effective field theory techniques. We emphasize the important role of jet substructure observables as they probe various aspects of the jet formation mechanism and allow us to study the medium properties in great details. The precise calculations require the systematic resummation and consistently including medium modifications. Specifically, we discuss the calculations of jet shapes and cross sections in proton-proton and lead-lead collisions at the LHC using soft-collinear effective theory, with Glauber gluon interactions in the medium. In the end we present the comparison between our calculations and the recent measurements at the LHC with very good agreement. We conclude that precise jet modification studies in heavy ion collisions has been initiated.}

\FullConference{Fourth Annual Large Hadron Collider Physics\\
		13-18 June 2016\\
		Lund, Sweden}

\usepackage{url,overcite}
\usepackage{multicol,psfrag,cite,cancel}
\usepackage[svgnames]{xcolor}

\newcommand{\be}{\begin{equation}}
\newcommand{\ee}{\end{equation}}
\newcommand{\bea}{\begin{eqnarray}}
\newcommand{\eea}{\end{eqnarray}}

\begin{document}

\section{Introduction}

The creation of the quark-gluon plasma (QGP) at the Relativistic Heavy Ion Collider (RHIC) and the Large Hadron Collider (LHC) has long been established with the observation of jet quenching -- the strong suppression of hadron and jet production cross sections in heavy ion collisions. Understanding the precise properties of this novel medium and QCD phases is at the heart of all experimental and theoretical efforts. To achieve the goal, various hard and soft probes of the QGP are employed. This talk focuses on the use of hard probes and, more specifically, jets which are produced and propagate within the medium.

Hard probes utilize theoretically well-controlled hard scattering processes and through the study of their modifications in heavy ion collisions infer the medium properties. As mentioned earlier, the first observables in this context are the production cross sections of energetic hadrons and jets. Their suppressions were understood using the parton energy loss picture with the fact that cross sections are steeply falling with the transverse momentum. However, it has been realized that the hadron and jet cross sections along are not sufficient to disentangle the detailed jet formation mechanisms in the medium, and we need more differential and correlated measurements of event-wide QCD radiations and jet substructures.

Jet substructure observables can resolve jets and the QGP at different energy scales. Therefore a series of jet substructure measurements sensitive to radiations in different energy regimes will allow us to study and test each stage of the parton shower evolution. On big advantage of studying jet substructures is that they are more sensitive to the final-state, jet-medium interactions, which allows us to more cleanly separate the initial-state, cold nuclear matter (CNM) effects. In this talk, I will take the jet shape \cite{Ellis:1992qq}, which probes the transverse energy distribution inside jets, as a classic example of jet substructure observables and discuss its precise calculation.

The fact that jet quenching is a multi-scale problem strongly motivates the use of effective field theory techniques in jet substructure calculations. The idea of effective field theory is to identify the dominant degrees of freedom, or QCD modes, which give leading contributions to the observable. Already for jets in proton-proton collisions we see that several hierarchical scales can be dynamically generated. This leads to large perturbative logarithms which need to be resummed, and effective field theory techniques provide a systematic way to perform resummation using renormalization-group evolutions. In the presence of the medium with the characteristic energy scale around the medium temperature, more low-energy scales reflect the collective behaviors and medium excitations. The proximity of the jet and medium scales causes interference in the jet-medium interactions which underlies the Landau-Pomeranchuk-Migdal (LPM) effect. The induced radiations and medium responses are altogether captured within jets, and effective field theory techniques are extremely useful in decomposing each component and simplifying the calculation.

I will present the jet shape calculation in both proton-proton \cite{Chien:2014nsa} and lead-lead \cite{Chien:2015hda} collisions using soft-collinear effective theory (SCET) \cite{Bauer:2000ew, Bauer:2000yr, Bauer:2001ct, Bauer:2001yt, Bauer:2002nz, Beneke:2002ph}, with Glauber gluon interactions in the medium \cite{Idilbi:2008vm, Ovanesyan:2011xy}. I will introduce the jet shape and discuss its factorization theorem in SCET. The resummation of the jet shape is performed at next-to-leading logarithmic (NLL) accuracy using renormalization-group techniques. In heavy ion collisions the medium modification to the jet shape as well as the jet energy loss are calculated using the medium-induced splitting functions \cite{Ovanesyan:2011kn, Fickinger:2013xwa}. In the end I will compare the theoretical calculation with the data from ALICE \cite{Abelev:2013kqa}, ATLAS \cite{Aad:2012vca,Aad:2014bxa} and CMS \cite{Chatrchyan:2013kwa,CMS:prelim} and make predictions for the upcoming Run 2 measurements.

\section{The jet shape}
\label{sec:obs}

\begin{figure}[top]
\center
\psfrag{x}{\footnotesize $r$}
\psfrag{z}{\footnotesize $\rho(r)$}
    \includegraphics[height=4cm, trim = 10mm -15mm 0mm 0mm]{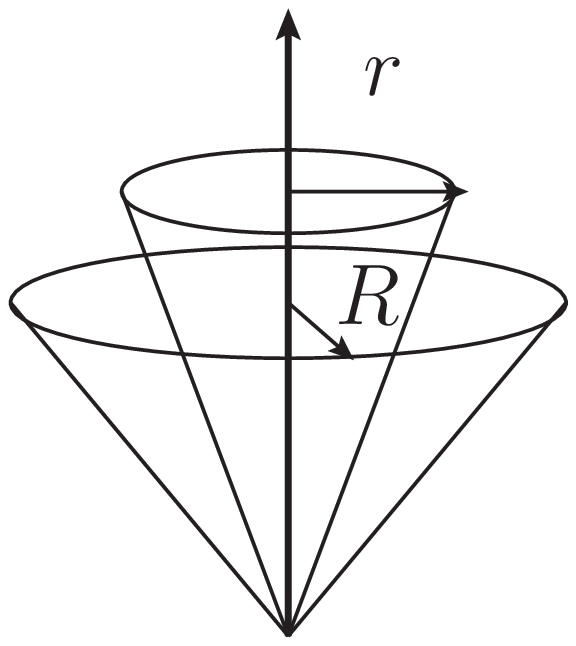}~~~~~~~~~~~~~~~
    \includegraphics[height=4cm, trim = 0mm 0mm 0mm 0mm]{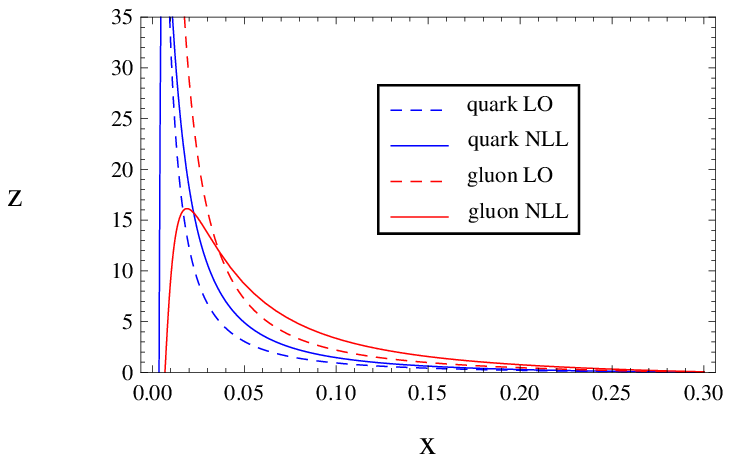}
\caption{Left: The jet shape probes the transverse energy distribution within a reconstructed jet as a function of the subcone size $r$. Right: The resummed (solid) and fixed-order (dashed) differential jet shapes of quark-initiated (blue) and gluon-initiated (red) jets. }
\end{figure}
The jet shape \cite{Ellis:1992qq} probes the transverse energy profile inside a reconstructed jet with radius $R$. It is defined as the fraction of the transverse energy $E_T$ of the jet within a subcone of size $r$ around the jet axis $\hat n$ (FIG. \ref{sec:obs}),
\be
    \Psi_J(r)=\frac{\sum_{i,~d_{i\hat n}<r} E^i_T}{\sum_{i,~d_{i\hat n}<R} E^i_T}\;.
\ee
Here $d_{i\hat n}=\sqrt{(\eta_i-\eta_{jet})^2+(\phi_i-\phi_{jet})^2}$ is the Euclidean distance between the $i$-th particle in the jet and the jet axis on the pseudorapidity-azimuthal angle $(\eta,\phi)$ plane. Note that by definition $r \leq R$ and $\Psi_J(R)=1$. This quantity is averaged over all jets and its derivative $\rho(r)$ describes how the transverse energy is differentially distributed in $r$.
\be
    \Psi(r)=\frac{1}{N_J}\sum_{J=1}^{N_J} \Psi_J(r),~~~~~~~~\rho(r)=\frac{d}{dr}\Psi(r)\;.
\ee
In heavy ion collisions, the modification of the jet shape is conventionally quantified by taking the ratio of the jet shapes in nucleus-nucleus and proton-proton collisions $\rho^{AA}(r)/\rho^{pp}(r)$.

When $r$ is parametrically smaller than $R$, already in proton-proton collisions the perturbative calculation of the jet shape suffers from the presence of large logarithms of the form $\alpha_s^n\log^m r/R~(m\leq 2n)$ which need to be resummed. These logarithmic terms come from the infrared structure of QCD which enhances the emissions of soft and collinear radiations. The right plot in FIG. \ref{sec:obs} shows the necessity of resummation in the jet shape calculation. As can be seen, the fixed-order predictions (dashed curves) diverge as $r$ goes to zero, and the effect of resummation (solid curves) is significant. There is no region of $r$ where the fixed-order calculation of the jet shape is sufficient. Also, the jet shapes of quark-initiated and gluon-initiated jets are different due to their different color charges. Qualitatively quark-initiated jets are more collimated whereas the energy within gluon-initiated jets are more spread out. In the following, I will first demonstrate how the all-order resummation of the jet shape is performed using SCET \cite{Bauer:2000ew, Bauer:2000yr, Bauer:2001ct, Bauer:2001yt, Bauer:2002nz, Beneke:2002ph}.

\section{The factorization theorem of the jet shape in SCET}
\label{sec:fac}

\begin{figure}[top]
\center
\includegraphics[width=0.26\linewidth]{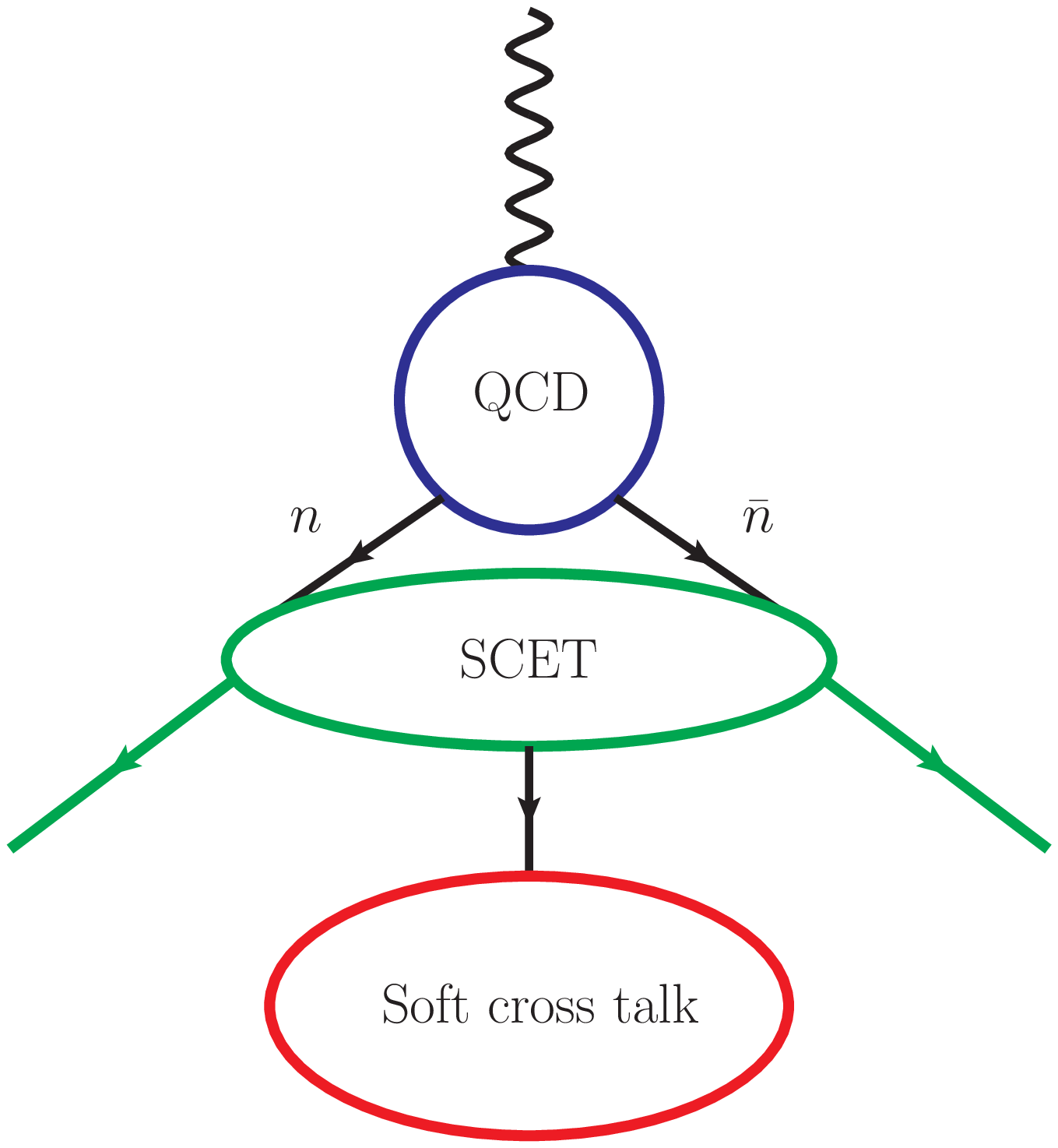}~~~~~~~~~~~~~~
\includegraphics[width=0.28\linewidth]{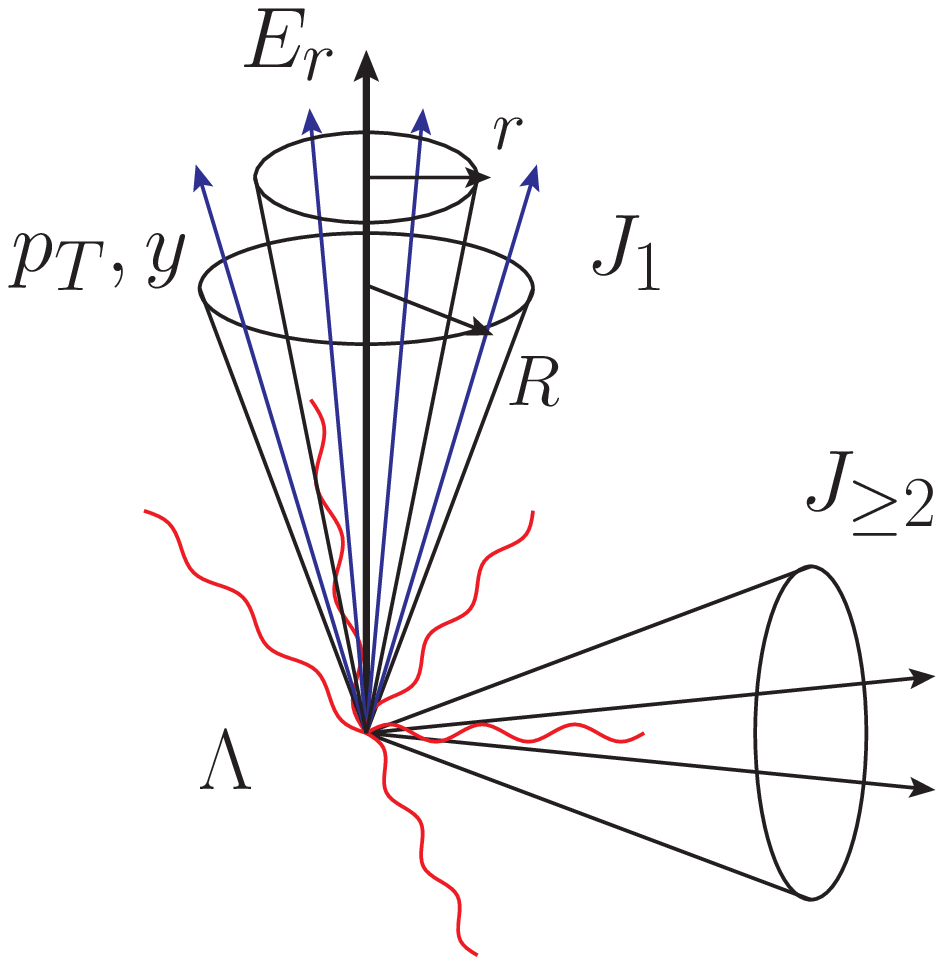}
\caption{Left: The factorization of the {\color{blue}hard}, {\color{green!50!black}collinear} and {\color{red}soft} sectors in SCET. Right: The schematic illustration of the jet production and the measurement of the jet shape.}
\label{fig:SCET}
\end{figure}

Effective field theory techniques are useful whenever there is clear scale separation. In events with the production of energetic and collimated jets, at least three distinct scales will emerge: the hard scale where the hard scattering processes happen, the jet scale which is the typical scale of the momentum transverse to the jet direction, and the soft scale which is the energy scale of the dominant soft radiation (FIG.\ref{fig:SCET}). A power counting parameter is defined as the ratio between hierarchical scales therefore is a small expansion parameter. SCET systematically power expands the QCD contributions, and the contributions from different sectors to a physical cross section are described by the hard, jet and soft functions respectively.

The key ingredient in SCET is the factorization of the hard, collinear and soft sectors at leading power. One first matches SCET to QCD at the hard scale by integrating out the contributions from the hard modes and these are encoded in the hard function. Then one identifies the collinear and soft modes which give the leading power contributions to the observable and are described by collinear and soft Wilson lines. Depending on the observables, the way collinear and soft modes contribute to the cross section may differ and these are encoded in the jet and soft functions. The hard, jet and soft functions can be calculated separately in each sector with a single characteristic scale \footnote{The soft function of exclusive observables may still be multi-scaled and need to be refactorized \cite{Ellis:2010rwa,Becher:2015hka,Chien:2015cka}.}, and the factorization theorem puts them together and constructs the physical cross section.

Without loss of generality we will demonstrate the factorization structure of the jet shape in $e^+e^-$ collisions (FIG. \ref{fig:SCET}). At hadron colliders the contributions from initial state radiation are power suppressed. The differential cross section of the production of $N$ jets with transverse momenta $p_{T_i}$ and rapidity $y_i$, the energy $E_r$ inside the cone of size $r$ in one jet, and an energy cutoff $\Lambda$ outside all the jets can be factorized into a product of the hard, jet and soft functions,
\be
    \frac{d\sigma}{dp_{T_i}dy_i dE_r}
    =H(p_{T_i},y_i,\mu)J_1^{\omega_1}(E_r,\mu) J_2^{\omega_2}(\mu)\dots J_N^{\omega_N}(\mu)S_{1,2,\dots N}(\Lambda,\mu).
\ee
$H(p_{T_i},y_i,\mu)$ is the hard function which is the square of the matching coefficient of SCET to QCD at the hard scale. $J^{\omega}(E_r,\mu)$ is the jet function which is the probability distribution of the energy $E_r$ inside a subcone of size $r$ in the jet with energy $\omega=2E_J$. All the other jet functions $J^{\omega}(\mu)$ are unmeasured jet functions \cite{Ellis:2010rwa} without measuring the substructure of the jet. $S_{1,2,\dots N}(\Lambda,\mu)$ is the soft function and it describes how soft radiation is constrained by the measurements. Detailed expressions of the operator definitions of the jet and soft functions can be found in \cite{Chien:2014nsa} and are not given here. Note that the factorization theorem simplifies dramatically and has a product form instead of a convolution. The contribution from the soft mode to the subcone energy $E_r$ is power suppressed. The hard, jet and soft functions as well as their anomalous dimensions can be calculated order by order in perturbative theory at their characteristic scales. Large logarithms of the ratio between hierarchical energy scales are resummed through the renormalization-group evolution of these functions.

The averaged energy inside the cone of size $r$ in jet 1 is the following,
\be
    \langle E_r\rangle_{\omega_1}
    =\frac{\int dE_r E_r\frac{d\sigma}{dp_{T_i}dy_idE_r}}{\frac{d\sigma}{dp_{T_i}dy_i}}
    =\frac{{\color{blue}H({p_T}_i,y_i,\mu)}J^{\omega_1}_{{E,r}_1}(\mu){\color{green!50!black}J^{\omega_2}_2(\mu)}\dots {\color{red}S_{1,2,\dots}(\Lambda,\mu)}}{{\color{blue}H({p_T}_i,y_i,\mu)}J^{\omega_1}_1(\mu){\color{green!50!black}J^{\omega_2}_2(\mu)}\dots {\color{red}S_{1,2,\dots}(\Lambda,\mu)}}
            =\frac{ J^{\omega_1}_{{E,r}_1}(\mu)}{J^{\omega_1}_1(\mu)}\;,
\ee
and $J_{E,r}^{\omega}(\mu)=\int dE_r E_r~J^\omega(E_r,\mu)$ is referred to as the jet energy function. Note the huge cancelation between the hard, unmeasured jet and soft functions, resulting in just the ratio between two jet functions associated with jet 1. This implies that the jet shape is insensitive to the underlying hard scattering process as well as the other part of the event when we neglect power corrections. It only depends on the energy and the identity of the jet-initiating parton (quark or gluon). In the end, the integral jet shape is weighted with the jet production cross sections with proper phase space cuts on $p_T$ and $y$,
\be
    \Psi(r)=\frac{1}{\sigma_{\rm total}}\sum_{i=q,g}\int_{PS} dp_Tdy \frac{d\sigma^{i}}{dp_Tdy}\Psi^i_\omega(r)\;,~{\rm where}~
    \Psi_\omega(r)
    =\frac{J_{E,r}(\mu)/J(\mu)}{J_{E,R}(\mu)/J(\mu)}
    =\frac{J_{E,r}(\mu)}{J_{E,R}(\mu)}\;.
\ee

\section{Scale hierarchy and renormalization-group evolution}
\label{sec:RG}

\begin{figure}[top]
\center
\includegraphics[height=3cm]{jetshape}~~~~~~
\includegraphics[height=3cm]{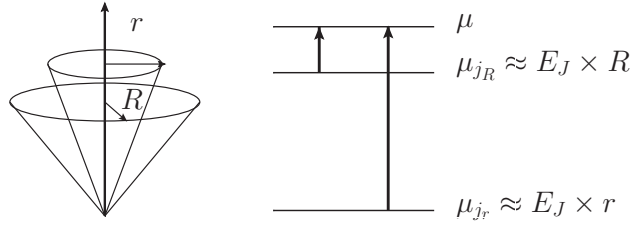}
\caption{The renormalization-group evolution of the jet energy function between $\mu_{j_r}$ and $\mu_{j_R}$ resums $\log \mu_{j_r}/\mu_{j_R}=\log r/R$.}
\label{fig:RG}
\end{figure}

The renormalization-group evolution of the jet energy function allows us to resum the jet shape. It satisfies the following RG equation ,
\be
    \frac{d J^{q}_{E,r}(r,R,\mu)}{d\ln\mu}
    =\left[-C_F~\Gamma_{\rm cusp}(\alpha_s)\ln\frac{\omega^2\tan^2\frac{R}{2}}{\mu^2}-2\gamma_{J^q}(\alpha_s)\right]J^{q}_{E,r}(r,R,\mu)\;,
\ee
for quark jets (for gluon jets with the color factor $C_A$), and $\Gamma_{\rm cusp}$ is the cusp anomalous dimension. Using the RG equation we can evolve the jet energy function from its natural scale $\mu_{j_r}$ to the renormalization scale $\mu$. Note that the integral jet shape $\Psi_\omega$ is renormalization group invariant,
\be
    \Psi_\omega=\frac{J_{E,r}(\mu)}{J_{E,R}(\mu)}=\frac{J_{E,r}(\mu_{j_r})}{J_{E,R}(\mu_{j_R})}U_J(\mu_{j_r},\mu_{j_R})\;.
\ee
The scale $\mu_{j_r}=\omega\tan\frac{r}{2}\approx E_J \times r$ can be identified \cite{Chien:2014nsa} which eliminates the large logarithms in the fixed-order calculation of the jet energy function $J_{E,r}(\mu_{j_r})$. The hierarchy between $r$ and $R$ induces two hierarchical jet scales $\mu_{j_r}$ and $\mu_{j_R}$, and $U_J(\mu_{j_r},\mu_{j_R})$ is the RG evolution kernel which resums the logarithms of the ratio between $\mu_{j_r}$ and $\mu_{j_R}$ (FIG.\ref{fig:RG}).

\section{Baseline jet shape calculations}
\label{sec:pp}

\begin{figure}[top]
\center
\psfrag{x}{$r$}
\psfrag{z}{$\rho(r)$}
\includegraphics[width=0.45\linewidth]{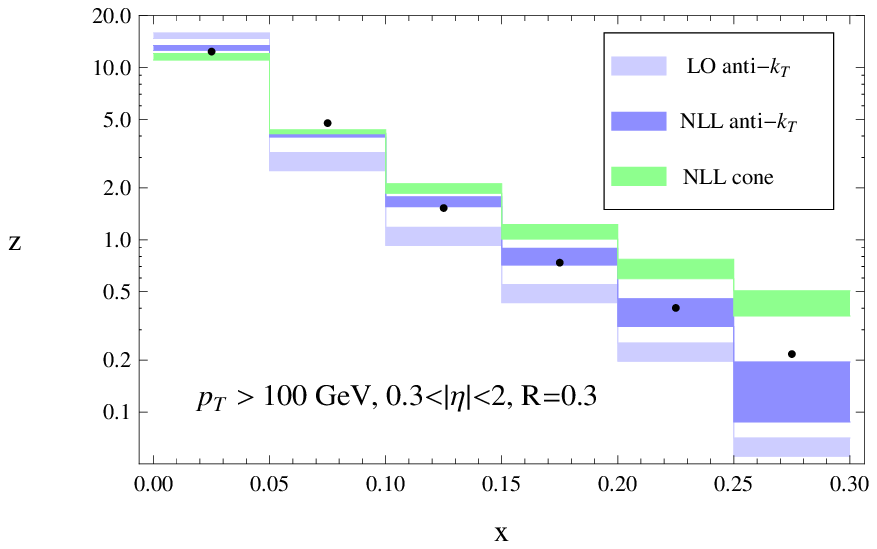}~~~~
\includegraphics[width=0.451\linewidth]{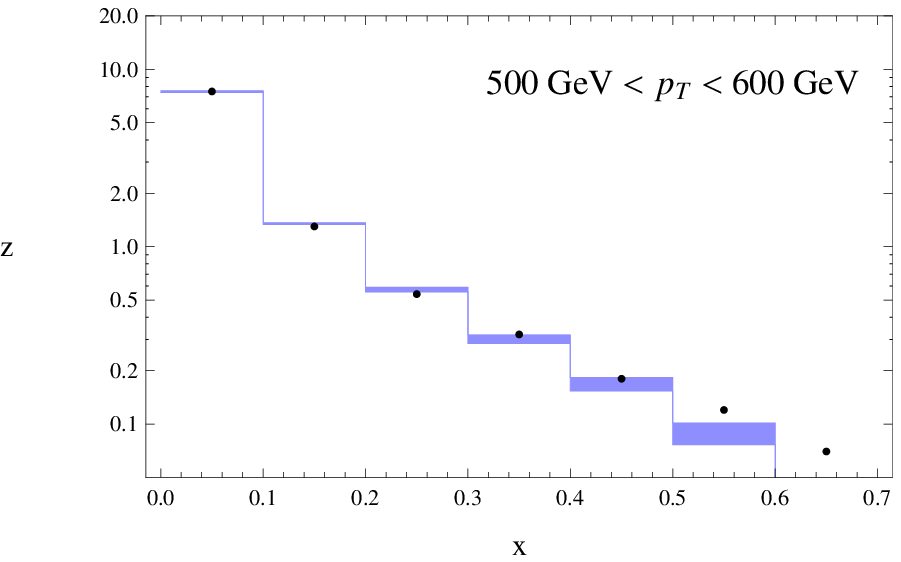}
\caption{Left: The differential jet shape for anti-$\rm k_T$ jets with $R=0.3$, $p_T>100$ GeV and $0.3<|y|<2$ in proton-proton collisions at the 2.76 TeV LHC. The black dots are the data from CMS. The shaded blue boxes are the LO (light) and NLL (dark) calculations, while the shaded green boxes are the NLL calculation for cone jets. Right: The differential jet shape for anti-$\rm k_T$ jets with $R=0.7$ at the 7 TeV LHC.}
\label{fig:resum}
\end{figure}

We compare our calculations with the CMS measurements at the 2.76 TeV and 7 TeV LHC (FIG.\ref{fig:resum}). Both the resummed and fixed-order results are shown to emphasize the importance of performing the resummation, with the NLL calculation agreeing much better with data than the fixed-order result. We also plot the jet shapes for jets reconstructed using the anti-$k_T$ and the cone algorithms to illustrate the jet algorithm dependence, and you can see the significant difference. The theoretical uncertainties are estimated by varying the jet scales $\mu_{j_r}$ and $\mu_{j_R}$ in the resummed expressions, shown as the shaded boxes in the plots. In the tail region ($r\approx R$) the power corrections become large which cause the discrepancy with data. The CMS collaboration performs a background subtraction which removes some of the power corrections, and our NLL calculation agrees with the data very well. This sets the baseline calculation and I now move on to discuss the medium modification of the jet shape.

\section{Multiple scattering and $\rm SCET$ with Glauber gluons}

\begin{figure}[top]
\center
\includegraphics[height=2.5cm, trim = 0mm 0mm 0mm 8mm]{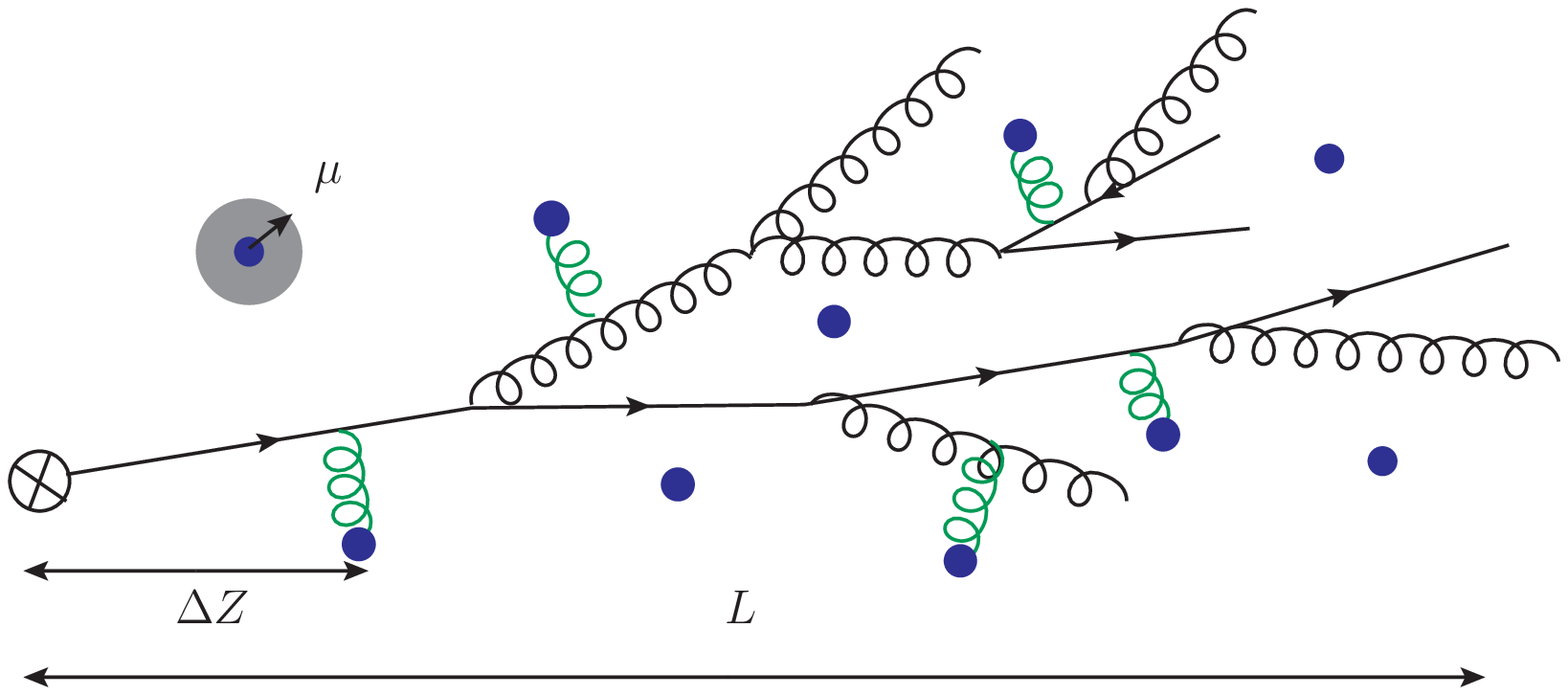}~~~
\includegraphics[height=4cm]{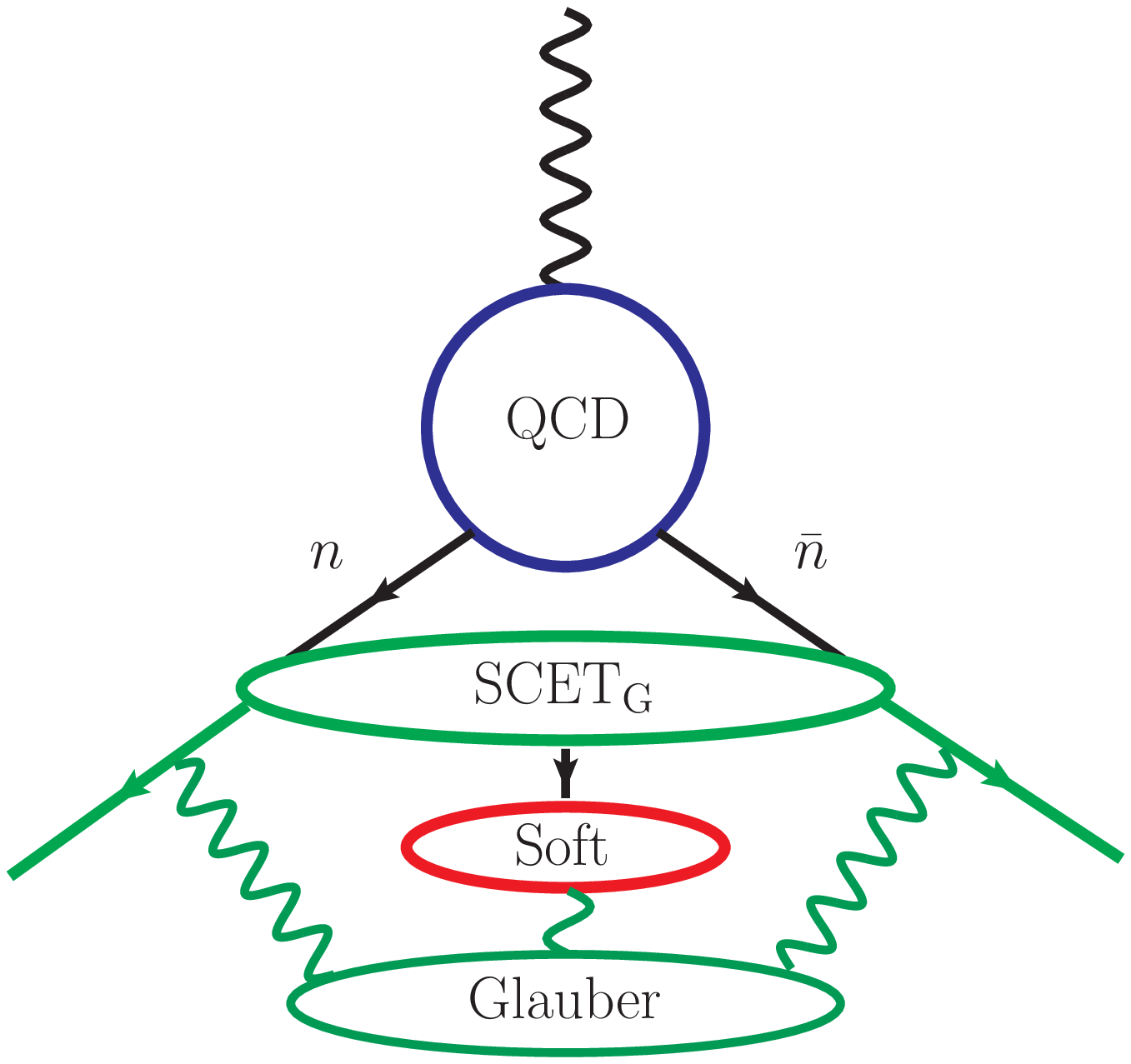}~~~
\includegraphics[height=2.5cm, trim = 0mm 0mm 0mm 2mm]{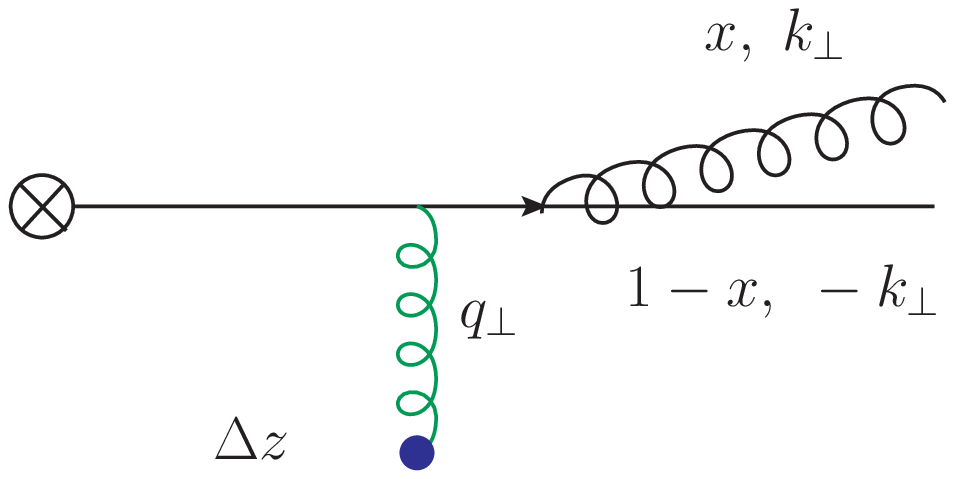}
\caption{Left: Illustration of the emergent scales in the medium jet formation. Middle: The factorization of the {\color{blue}hard}, {\color{green!50!black}collinear}, {\color{red}soft} and {\color{green!50!black}Glauber} sectors in $\rm SCET_G$. Right: Parton splitting kinematics at leading order.}
\label{fig:medium}
\end{figure}

Coherent multiple scattering and the induced bremsstrahlung are the qualitatively new features of the in-medium parton shower evolution (FIG.\ref{fig:medium}, left). In the presence of the medium, more characteristic scales emerge throughout the jet formation. The Debye screening scale $\mu$ sets the range of interaction, and the parton mean free path $\lambda$ determines the significance of multiple scattering in the medium. On the other hand, the radiation formation time $\tau$ sets the scale where the jet and the medium can be resolved, and it is closely related to the parton splitting kinematics. The interplay among these characteristic scales can result in different interference patterns. When $\tau$ is much smaller than $\lambda$, this is the Bethe-Heitler incoherent limit where multiple scattering are uncorrelated. On the other hand, when $\tau$ is much larger than $\lambda$, there is destructive interference among multiple scattering which underlies the Landau-Pomeranchuk-Migdal effect.

With the above physical picture of jet quenching with multiple emergent scales, we would like to exploit effective field theory techniques by extending SCET with the jet-medium interaction \cite{Idilbi:2008vm, Ovanesyan:2011xy}. We identify that the Glauber gluon is the relevant mode describing the momentum transfer transverse to the jet direction between the jet and the medium, and the extended effective theory is dubbed $\rm SCET_G$ (FIG.\ref{fig:medium}, middle). The Glauber gluons are generated from the color charges in the medium. Given a medium model, we can consistently couple the medium to jets using $\rm SCET_G$. From thermal field theory and lattice QCD calculations, an ensemble of quasi particles with Debye screened potential and thermal masses is a reasonably valid parameterization of the medium properties. In this work we adopt this medium model with the Bjorken-expanded hydrodynamic evolution \cite{Bjorken:1982tu}.

\section{Medium-induced splitting function and Landau-Pomeranchuk-Migdal effect}

The key ingredients which enter the calculations of the medium modification of jet substructure observables are the medium-induced splitting functions \cite{Ovanesyan:2011kn, Fickinger:2013xwa}. At leading order, for a parton emitting an induced radiation with momentum fraction $x$ and transverse momentum $k_\perp$ (FIG.\ref{fig:medium}, right), the full splitting function is calculated and the result can be found in \cite{Ovanesyan:2011kn}. To illustrate the LPM effect, in the soft gluon limit the result reduces to the following expression,
\be
    \frac{dN^{med}_{q\rightarrow qg}}{dxd^2k_\perp}
    =\frac{C_F\alpha_s}{\pi^2}\frac{1}{x}\int_0^L\frac{d\Delta z}{\lambda}\int d^2q_\perp \frac{1}{\sigma_{el}}\frac{d\sigma_{el}}{d^2q_\perp}\frac{2k_\perp\cdot q_\perp}{k_\perp^2(q_\perp-k_\perp)^2}
    \textcolor{blue}{\Big[1-\cos\Big(\frac{(q_\perp-k_\perp)^2\Delta z}{x \omega}\Big)\Big]}\;,
\ee
with the effective cross section $\frac{1}{\sigma_{el}}\frac{d\sigma_{el}}{d^2q_\perp}=\frac{\mu^2}{\pi(q_\perp^2+\mu^2)^2}$. Here you can see the interference, cosine term, with its argument the ratio between the radiation formation time $\tau = \frac{x~\omega}{(q_\perp-k_\perp)^2}$ and the distance between scattering $\Delta z$. In the limit where $k_\perp$ goes to zero, such induced-splitting probability vanishes which is the LPM effect. The large angle bremsstrahlung takes away energy, resulting in the jet energy loss and the modification of the jet shape.

\section{Jet shape in heavy ion collisions}

\begin{figure}[top]
\center
\psfrag{x}{$r$}
\psfrag{w}{\small $\frac{\psi(r)^{\rm Pb}}{\psi(r)^{\rm P}}$}
    \includegraphics[height=4cm]{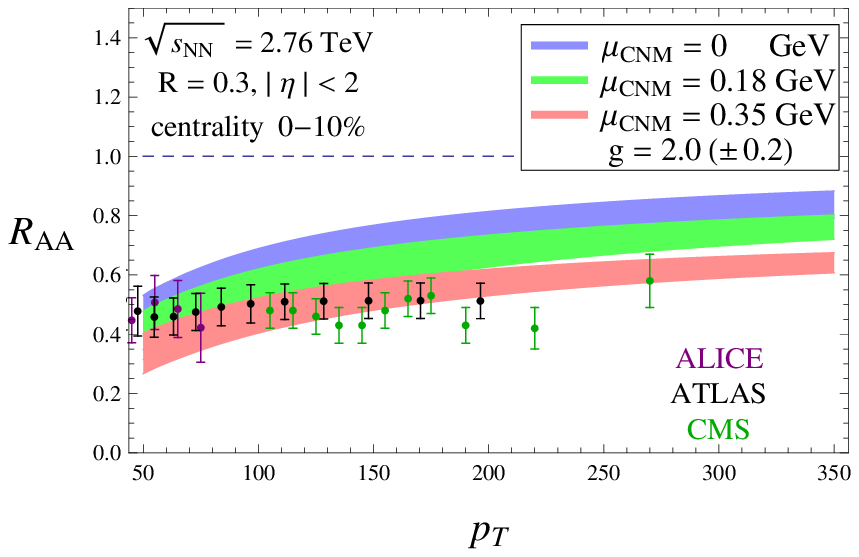}~~~~~
    \includegraphics[height=4.2cm]{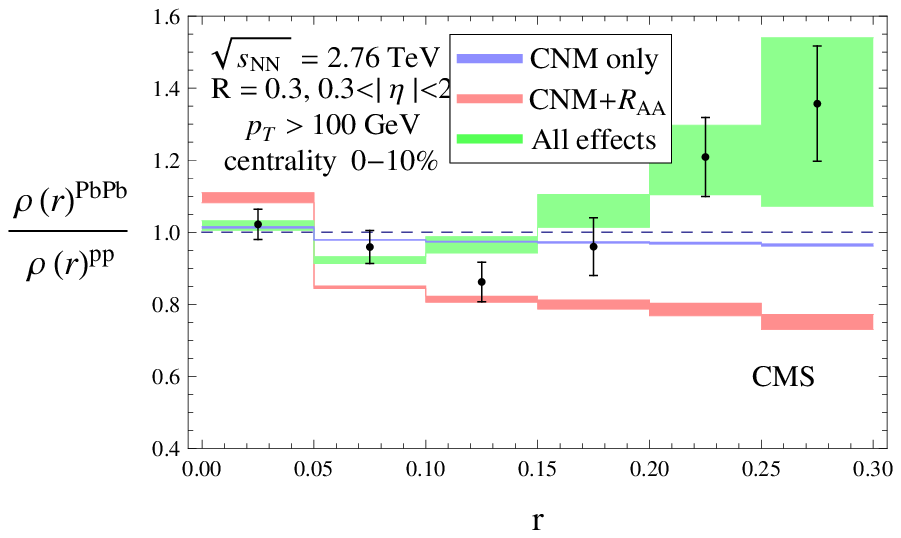}
\caption{Left: Nuclear modification factor $R_{AA}$ for anti-$k_T$ jets with $R=0.3$, $|\eta|<2$ as a function of $p_T$ in central lead-lead collisions at $\sqrt{s_{\rm NN}}=2.76$ TeV, with different cold nuclear matter effects implemented. Right: The ratio of the jet shapes in pp and PbPb collisions, with the CNM effect (blue), plus the cross section suppression (red), plus the jet-by-jet shape modification (green).}
\label{fig:result1}
\end{figure}

Due to its collinear nature, the jet shape can be calculated using the collinear parton splitting functions. At leading order,
\be
    J^{i}_{E,r}(\mu)=\sum_{j,k}\int_{PS} dxdk_\perp\Big[\frac{dN^{vac}_{i\rightarrow jk}}{dxd^2k_\perp}+{\color{blue}\frac{dN^{med}_{i\rightarrow jk}}{dxd^2k_\perp}}\Big] E_r(x,k_\perp)\;.
\ee
It is then straightforward to calculate the integral jet shape in heavy ion collisions.
\be
    \Psi(r)
    =\frac{J^{vac}_{E,r}+J^{med}_{E,r}}{J^{vac}_{E,R}+J^{med}_{E,R}}
    =\frac{\textcolor{red}{\Psi^{vac}(r)}J^{vac}_{E,R}+\textcolor{green!50!black}{J^{med}_{E,r}}}{J^{vac}_{E,R}+J^{med}_{E,R}}\;.
\ee
Because of the LPM effect, $J_{E,r}^{med}(r)$ contributes as a power correction without large logarithms. There is no extra soft-collinear divergence and the RG evolution of the jet energy function is the same as in vacuum. The jet shape is then averaged with the jet cross section which is significantly suppressed due to the jet energy loss in heavy ion collisions \cite{Chien:2015hda}. Note that gluon-initiated jets are quenched more than quark-initiated jets therefore the quark jet fraction is increased.

\begin{figure}[top]
    \psfrag{x}{$r$}
    \psfrag{w}{\small $\frac{\psi(r)^{\rm Pb}}{\psi(r)^{\rm P}}$}
        \includegraphics[height=3.0cm]{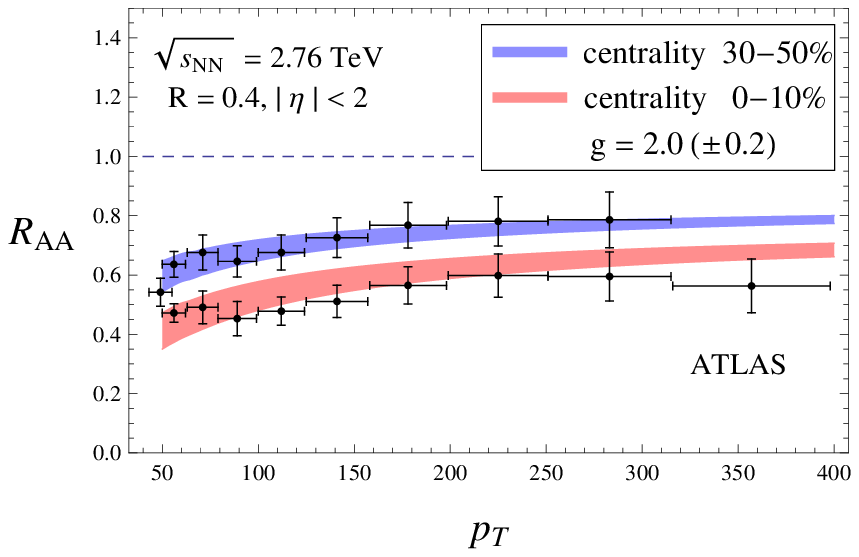}
        \includegraphics[height=3.1cm]{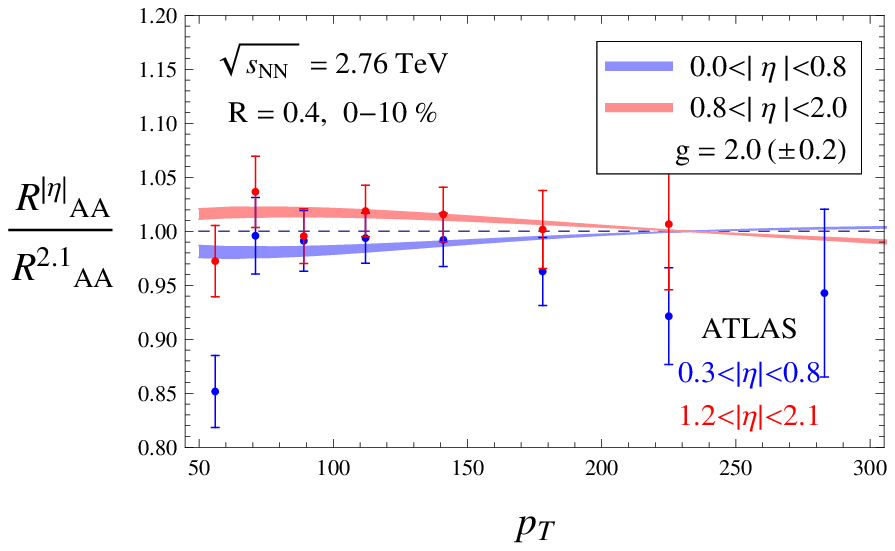}
        \includegraphics[height=3.1cm]{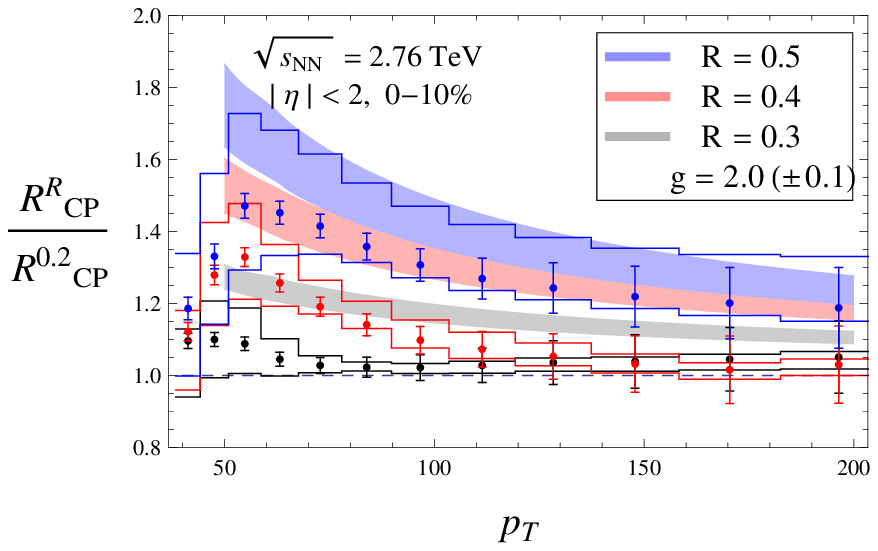}
\caption{Examination of the centrality (left), the jet rapidity (middle) and the jet radius (right) dependence in the jet cross section suppression. The theory calculations agree with the data very well.}
\label{fig:result2}
\end{figure}

\begin{figure}[t]
\psfrag{x}{$r$}
    \psfrag{w}{\small $\frac{\psi(r)^{\rm Pb}}{\psi(r)^{\rm P}}$}
        \includegraphics[height=3.1cm]{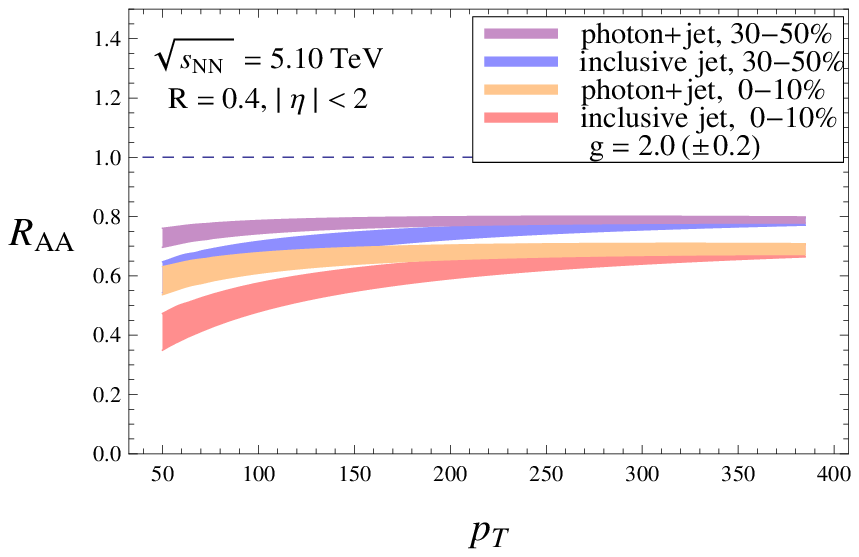}
        \includegraphics[height=3.1cm]{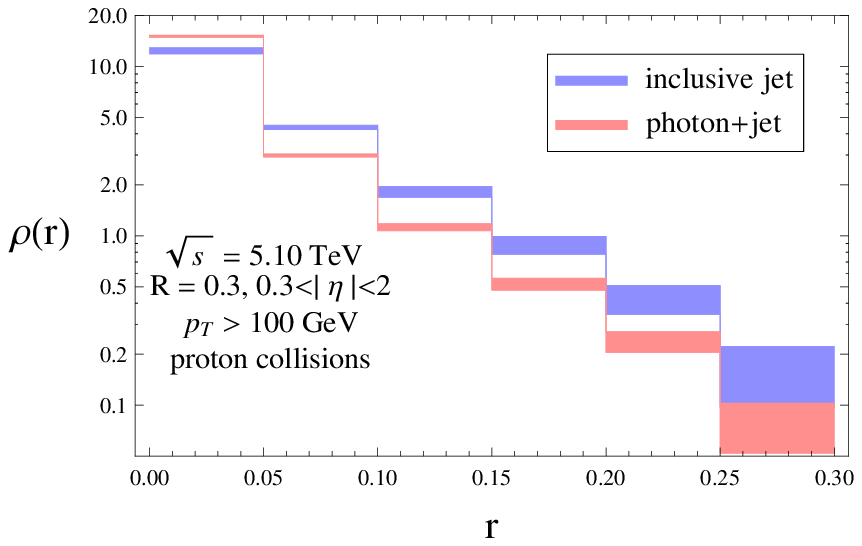}
        \includegraphics[height=3.1cm]{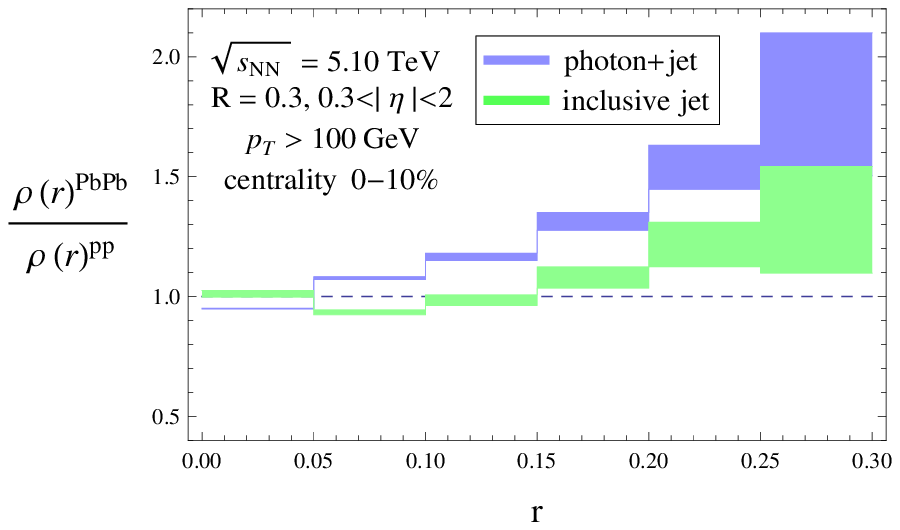}
\caption{Predictions for the jet cross section (left) and the jet shape (middle and right) at the $\sqrt{s_{\rm NN}}=5$ TeV LHC for inclusive and photon-tagged jets. Due to the higher quark jet fraction, for photon-tagged jets the cross section suppression is less and and the jet broadening is more manifest.}
\label{fig:result3}
\end{figure}

We present the theoretical calculations of the nuclear modification factor $R_{AA}$ for jet cross sections as well as the ratio of the differential jet shapes in lead-lead versus proton-proton collisions at $\sqrt{s_{\rm NN}}=2.76$ TeV (FIG.\ref{fig:result1}). The left plot shows the results with different CNM effects implemented in the calculations, and we compare with the measurements from ALICE \cite{Abelev:2013kqa}, ATLAS \cite{Aad:2012vca} and CMS \cite{CMS:prelim}. The theoretical uncertainty is estimated by varying the jet-medium coupling $g$. We see that the jet cross section is quite sensitive to the CNM effect. The right plot shows the jet shape calculation and studies its sensitivity to the CNM effect, the quark/gluon jet fraction and the jet-by-jet shape modification. The shaded boxes represent the scale uncertainty. We see that the non-trivial jet shape modification pattern observed at CMS \cite{Chatrchyan:2013kwa} is due to both the increase of the quark jet fraction, which tends to make the jet shape narrower, as well as the broadening of the jet-by-jet shape, and it is insensitive to the CNM effect. We also examine the dependence of $R_{AA}$ on centrality, the jet rapidity and the jet radius (FIG.\ref{fig:result2}) \cite{Aad:2012vca,Aad:2014bxa} and make predictions for the jet shape and cross section of inclusive and photon-tagged jets at the LHC Run 2 (FIG.\ref{fig:result3}).

\section{Conclusions}

The jet shape and cross section in proton and heavy ion collisions are calculated using the SCET extended with Glauber gluon interactions. The jet shape is resummed at NLL accuracy using renormalization-group techniques, and the baseline calculation is established. The medium modification to the jet shape is calculated using the medium-induced splitting functions. We find good agreement between our calculations and the data. The new understanding we gain is that the non-trivial jet shape modification pattern is due to the combination of the cross section suppression and the jet-by-jet broadening. The precise jet substructure studies using effective field theory techniques will continue to teach us qualitatively new features in the jet quenching phenomenology.

\section{Acknowledgments}

This work is supported by the US Department of Energy, Office of Science.

\end{document}